# GAW, Gamma Air Watch – A Large Field of View Imaging Atmospheric Cherenkov Telescope

M.C. Maccarone[a], G. Agnetta[a], P. Assis[b], B. Biondo[a], P. Brogueira[b], O. Catalano[a],
F. Celi[a], J. Costa[b], G. Cusumano[a], C. Delgado[c], G. Di Cocco[d], M.C. Espirito Santo[b],
P. Galeotti[f], S. Giarrusso[a], A. La Barbera[a], G. La Rosa[a], A. Mangano[a], T. Mineo[a],
M. Moles[e], M. Pimenta[b], F. Prada[e], F. Russo[a], B. Sacco[a], M.A. Sanchez[e],
A. Segreto[a], B. Tomé[b], A. de Ugarte Postigo[e], P. Vallania[g], C. Vigorito[f]

*(a) IASF-Pa/INAF, Istituto di Astrofisica Spaziale e Fisica Cosmica, Palermo, Italy*
*(b) LIP, Laboratório de Instrumentação e Física Experimental de Partículas, Lisbon, Portugal*
*(c) IAC, Instituto de Astrofisica de Canarias, Tenerife, Spain*
*(d) IASF-Bo/INAF, Istituto di Astrofisica Spaziale e Fisica Cosmica, Bologna, Italy*
*(e) IAA, Instituto de Astrofisica de Andalucia (CSIC), Granada, Spain*
*(f) Dept. of Physics, University of Turin, Torino, Italy*
*(g) IFSI-To/INAF, Istituto di Fisica dello Spazio Interplanetario, Torino, Italy*

Presenter: M.C. Maccarone (Cettina.Maccarone@pa.iasf.cnr.it), ita-cusumano-G-abs1-og27-poster

GAW, acronym for Gamma Air Watch, is a "path-finder" experiment to test the feasibility of a new generation of imaging atmospheric Cherenkov telescopes that join high flux sensitivity with large field of view capability. GAW is conceived as an array of three identical imaging telescopes disposed at the vertexes of an equilateral triangle, ~80 m side. Two main features characterize GAW with respect to all the existing and presently planned ground-based Cherenkov telescopes. The first difference concerns the optics system: GAW uses a Fresnel refractive lens (Ø 2.13 m) as light collector instead of classical reflective mirror. The second main difference is the detection working mode used: the detector at the focal surface operates in single photoelectron counting mode instead of the usual charge integration one.
The GAW array is planned to be located at the Calar Alto Observatory site, Spain, 2150 m a.s.l. During its first phase, only 6°×6° of the focal plane detector will be implemented; moving it along the field of view, the sensitivity of the telescopes will be tested observing the Crab Nebula with on-axis and off-axis pointing up to 20° and with energy threshold of 300 GeV and energy peak of 700 GeV. As path-finder, GAW will also monitor the Very High Energy activity of some flaring Blazars as well as will follow-up GLAST detections at high energies. In a second phase, the focal plane will be enlarged to cover a field of view of ±12°; pointing along different North-South directions, GAW would reach a survey of 360°×60° region of the sky.
GAW is a collaboration effort of Research Institutes in Italy, Portugal and Spain.

## 1. Introduction

Despite its youngness, Very High Energy (VHE) gamma-ray astronomy is considered a legitimate astronomical discipline with established sources, steady and variable, galactic and extragalactic, that, although limited in number, provide deep implications in the theoretical models. Presently, the main objectives of the ground-based VHE gamma-ray experiments concern the extension of the observation range towards lower energies (few tenths of GeV) so to meet and overlap the energy window proper of the gamma-ray satellite experiments, and the improvement of the flux sensitivity above 100 GeV. A lower energy range is being exploited with arrays of solar heliostats as STACEE, Solar-2, CELESTE, or GRAAL; moreover, the telescope MAGIC, with its 17 m aperture, is going to achieve a similar low energy threshold using the Imaging Atmospheric Cherenkov Technique (IACT). The second objective, i.e. reaching higher sensitivity in the region above 100 GeV, are achieved using arrays of telescopes as VERITAS, HESS, or CANGAROO III, which, thanks to their stereoscopic observational approach, furtherly increase the sensitivity capability of IACT telescopes.

Nevertheless, astronomical events can occur at unknown locations and/or random in time, and a large field of view (FoV) is then needed to increase their detection probability. To collect light, the present IACT telescopes use large mirror reflectors characterized by FoV of the order of few degrees; they cannot reach larger FoV due to the mirror optical aberrations, rapidly increasing with off-axis angles, and to the shadow of a larger detector onto the reflector. A large FoV is also a basic requirement to perform sensitive survey of the Galactic Plane as well as an estimation of the celestial gamma-ray diffuse emission. Such a survey cannot be easily performed by the ground-based IACT telescopes: as a result of their reduced FoV, such telescopes can survey the sky only if long exposure times are considered.

To overcome such limitations, an alternative solution could come from the use of refractive optics, as the Fresnel lenses: they can achieve large FoV; they maintain imaging stability against deformation; there is no central obstruction of the detector at the focal surface.

GAW, acronym for Gamma Air Watch, is a "path-finder" experiment, currently under definition, to test the feasibility of a new generation of IACT telescopes that join high flux sensitivity with large field of view capability using Fresnel lens, stereoscopic observational approach, and single photoelectron counting mode.

## 2. The Experiment

GAW is conceived as an array of three identical IACT telescopes disposed at the vertexes of an equilateral triangle. A detailed description of GAW is given in [1]; here we report two specific main features of the array as R&D experiment for a new generation Cherenkov telescopes.

A refractive optical system characterizes GAW: its ligth collector is a non commercial Fresnel lens (Ø 2.13 m) with focal length of 2.56 m and standard thickness of 3.2 mm. The lens material is UltraViolet (UV) transmitting acrylic with a nominal transmittance of ~95% from 330 nm to the near InfraRed; this material joins high transmittance and small refraction index derivative at low wavelength, reducing chromatic aberration effects. The lens design is optimized to have, at the wavelength of maximum intensity of the Cherenkov light convolved with the detector response (~360 nm) a quite uniform spatial resolution up to 30° (full angle) suitable to the requirements of the Cherenkov imaging. The baseline optics module for the GAW prototype is a single-sided, flat Fresnel lens optimized for a ±12° field of view. The flat lens is composed by a central core (Ø 50.8 cm) surrounded by a corona of petals extending for 40.6 cm out from the radius of the core, and a second outer corona of petals extending for 40.6 cm more. The central core will be made with constant depth aspheric grooves; the petals will have constant width aspheric grooves. The assembled lens will be constituted by one single piece for the central core, 12 pieces for the first petals ring, and 20 pieces for the outer petals corona. A spider support will maintain all the pieces together. The project of the optical system is a joint effort of the IASF/INAF Institute in Palermo and of the Fresnel Technologies, Fort Worth, Texas, which will manufacture the lens.

The second main feature characterizing GAW is the detector working mode used. The focal surface detector of each telescope consists of a grid of MultiAnode PhotoMultiplier Tubes (MAPMT) manufactured by Hamamatsu, series R7600-03-M64; the number of active channels (order of $10^4$) forming the detector at the focal surface makes it basically a large UV sensitive digital camera with high resolution imaging capability. The large dead area of the chosen MAPMT induces a low geometrical efficiency factor of ~50% on photon detection; in order to correct that, each MAPMT pixel is coupled to a Light Guide which allows to uniformly cover the FoV with a 80% average absorption (due to the Light Guide). The specific feature is that the GAW electronics design is based on single photoelectron counting mode (front-end), instead of the charge integration method widely used in the IACT telescopes, and on free-running method (data taking and read-out). The single photoelectron counting mode method is a well-established technique and it is used to measure the number of output pulses from the photosensors corresponding to incident photons. Small pixel size is required to minimize the probability of photoelectrons pile-up within intervals shorter than the given sampling time of 10 ns (GTU, Gate Time Unit). In such working mode the electronics noise and the MAPMT gain differences are kept negligible allowing lowering the energy threshold in spite of the relatively small dimension of the GAW light collector. The free-running method makes use of cyclic memories to continuously store system and ancillary data at a predetermined sampling rate; once a specialized trigger stops the sampling procedure, data are recovered from the memories and ready to be transferred to a mass memory.

Therefore, the single photoelectron counting mode, together with the stereoscopic observational approach, will guarantee an energy threshold of the order of few hundreds of GeV in spite of the relatively small dimension of the lens.

GAW is planned to be located at the Calar Alto Observatory (Sierra de Los Filabres, Almeria, Spain, 2150 m a.s.l.). Two phases are foreseen for the project:

- Firstly, only part of the GAW focal detector will be implemented to cover a FoV of 6°× 6°; the detector will be mounted on a rack frame allowing shifting it along the full size of the focal surface. Under this configuration, with energy threshold of 300 GeV and energy peak of 700 GeV, the sensitivity of the telescopes will be tested observing the Crab Nebula with on-axis and off-axis pointing up to 20°. GAW will also monitor the VHE activity of some flaring Blazars.

- In a second phase, once the feasibility of the method proposed has been proved, the focal plane detector will be enlarged to cover a FoV of ±12°. We plan to survey a region of 360°×60° of sky pointing along different North-South directions.

To evaluate the GAW expected performance, a complete end-to-end simulation chain has been developed, starting from the physical process to the event reconstruction and analysis, through the effect of the atmospheric absorption and the detector response. The first step of the simulation chain, performed by using the CORSIKA code [2], mainly concerns the generation of Cherenkov light, at level of single photons, associated to air showers induced by different primaries; moreover, it includes the effects of the atmospheric absorption and a set of detector parameters (optics transmission, MAPMT photoelectron efficiency, …) values of which are considered nominal at this stage of the project (GAW in reduced configuration). This step is computationally time-consuming and a library of more than 6000 simulated air showers (protons and gamma) have been produced in the energy range 0.3-30 TeV as well as Crab-like spectrum and mono-energetic primaries, on-axis and off-axis.

The Cherenkov files are then analyzed by a proper user-defined homemade code that includes other peculiarities of the detector which can be easily updated whenever the hardware status of the telescope is modified. Among these, the GAW geometrical configuration, the optics size and its spread function, the effect of the light guides, the remaining efficiency factors, as well as the trigger electronics and the average expected value of light diffuse background are included. A threshold of 14 photoelectrons per event in each telescope, together with the coincidence in the other two telescopes is applied; this choice reduces the fake trigger rate due to the diffuse background to a negligible level.

A first group of simulation has been generated considering the three telescopes located at the vertices of a triangle 80 m side (Fig. 1 shows a possible accommodation of the GAW array at Calar Alto Observatory site). The fiducial area taken into account was a square of 1520×1520 m$^2$ around the triangle center; cores of the Cosmic Ray (gamma and protons) events were randomly distributed in this large square. As in the standard analysis of stereoscopic observations, the reconstruction of the arrival direction is obtained superimposing the Cherenkov images of the three telescopes in a common plane and determining the point of minimal distance among the major axes of the image ellipses (as sketched in Fig.2). The direction of each major axis is determined by minimizing its distance with respect to the photoelectrons position after suitable suppression of the more external points, mainly due to the diffuse light background.

The analysis of GAW simulated data in the range 0.3-30 TeV is in progress; early results on the expected performances are encouraging and show that GAW can operate at low energy threshold with deep flux sensitivity comparable to the present telescopes (details will be presented at the time of this Conference).

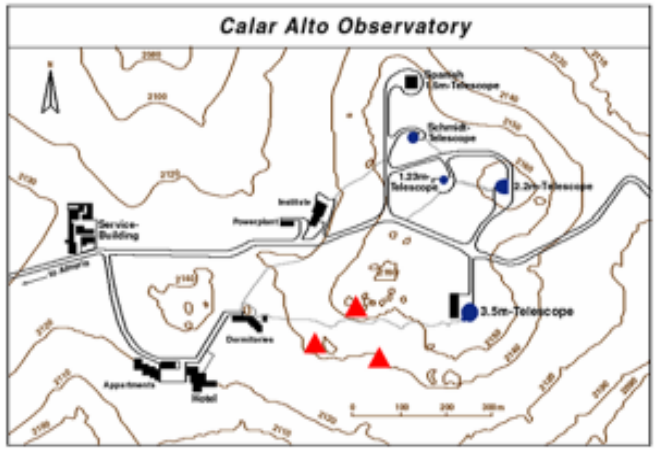
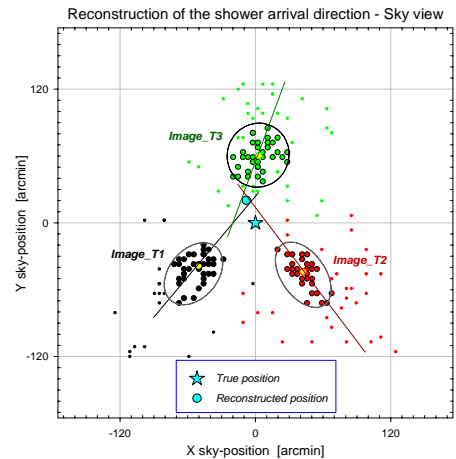

**Figure 1.** GAW array (red filled triangles) at Calar Alto site.

**Figure 2.** Example of sky-reconstruction of the shower arrival direction.

## 3. Conclusions

GAW is and R&D and path-finder experiment to test the possibility of a new generation of IACT telescopes that join high flux sensitivity with large field of view capability using Fresnel lens, single photon counting mode, and stereoscopic observational approach. Thanks to the single photon counting mode, which allows to operate with a very low photoelectrons threshold (~1/8 respect to the value required by the present IACTs), an enlargement of the GAW optics diameter could allow a very deep sensitivity observing a field of view one hundred times larger than what the present IACT telescopes can reach.